# Aluminum and Gallium Distribution in the $Lu_3(Al_{5-x}Ga_x)O_{12}$:Ce Multicomponent Garnet Scintillators Investigated by the Solid-State NMR and DFT calculations

Yu. Zagorodniy[1,2], V. Chlan[1], H. Štěpánková[1], Y. Fomichov[1,2], J. Pejchal[3], V. Laguta[2,3], M. Nikl[3]

[1]*Charles University, Faculty of Mathematics and Physics, V Holešovičkach 2, 180 00 Prague 8, Czech Republic*
[2]*Institute for Problems of Materials Science NAS Ukraine, Krjijanovsky 3, 03142 Kyiv, Ukraine*
[3]*Institute of Physics, Academy of Sciences of the Czech Republic, Cukrovarnicka 10, 162 00 Prague 6, Czech Republic*

**Abstract**

Distribution of aluminum and gallium atoms over the tetrahedral and octahedral sites in the garnet structure was studied in the mixed $Lu_3Al_{5-x}Ga_xO_{12}$ crystals using the $^{27}Al$ and $^{71}Ga$ MAS NMR together with the single crystal $^{71}Ga$ NMR. The experimental study was accompanied by theoretical calculations based on the density functional theory in order to predict the tendency in substitutions of Al by Ga in the mixed garnets. Both experimental and theoretic results show a non-uniform distribution of Al and Ga over the tetrahedral and octahedral sites in the garnet structure, with strong preferences for Ga, having larger ionic radius than Al, to occupy the tetrahedral site with smaller volume in the garnet structure. The quadrupole coupling constants and chemical shift parameters for Al and Ga nuclei have been determined for all the studied compounds as well as electric field gradients at Al and Ga nuclei were calculated in the framework of the density functional theory.

## 1. Introduction

$Lu_3Al_{5-x}Ga_xO_{12}$ multicomponent garnets doped with Ce belong to the group of functional materials of the general formula $(A1,A2)_3(B1,B2)_5O_{12}$ (where A1,A2 = Y, Lu, Gd, La and B1,B2 = Al, Ga, Sc), which, being doped with Nd, Ce, Pr, etc. are widely used nowadays as an infrared laser medium,[1] blue-to-yellow downconverter phosphor in white light emitting diodes,[2,3] and very fast and efficient scintillators with high light yield above 50000 photon/MeV[4-7] thus approaching to the theoretical limit expected for the garnet host.[8]

In recent years, much efforts were undertaken to further improve the scintillation performance of the $Lu_3Al_5O_{12}$ (LuAG) based crystals by band gap engineering and by changing the energy position of localized $Ce^{3+}$ electronic levels within the host band gap by choosing appropriate composition of the multicomponent compound (see e.g. Refs. 9-12). In particular, it was shown that Ga substitution for Al in the garnet crystals leads to the shift of the conduction band edge to lower energies leaving the top of the valence band almost unchanged. It results in the elimination of shallow electron traps (arising predominantly from cation antisite defects, e.g., $Lu^{3+}$ on $Al^{3+}$ sites and vice versa) due to location of their electronic levels within the conduction band. This has been shown both experimentally (see, e.g. Refs.12,13) and theoretically.[14] The decrease of the band gap in the mixed crystals is expected from the increase of the unit cell volume as the larger Ga atoms substitute the smaller Al atoms (0.535 Å and 0.62 Å for VI-fold $Al^{3+}$ and $Ga^{3+}$; and 0.39 Å and 0.47 Å for IV-fold $Al^{3+}$ and $Ga^{3+}$, respectively).[15] However, the selective shift of the conduction band edge is an exclusive property of the Ga ion associated with its electronic structure.

The second effect expected from the substitution of Al by Ga in the mixed crystals of $Lu_3Al_{5-x}Ga_xO_{12}$ is the change in the distribution of Al over the tetrahedral and octahedral sites in the $O_h^{10} - Ia3d$ garnet lattice.[16] It is known from the previous studies of the classical $Y_3Al_5O_{12}$ garnet material (e.g. Refs. 17-20) that Ga ions prefer to occupy the tetrahedral positions so that the occupation numbers Al(IV)/Al(VI) and Ga(IV)/Al(VI), which is equal to 3/2 in both $Y_3Al_5O_{12}$ and $Y_3Ga_5O_{12}$, essentially deviates from 3/2 in the case of $Y_3Al_{5-x}Ga_xO_{12}$ solid solutio.[17,20] The substitutions of Al for Ga and vice versa in the octahedral or tetrahedral positions obviously have a different effect on the electronic band structure of the garnet host material. More, as it was theoretically shown,[21] positioning of Ga in the octahedral or tetrahedral sites results in different changes of lattice parameters. Therefore, the knowledge of the cation distribution over the tetrahedral and octahedral sites is an important task in predictable engineering of the optical and scintillation properties of mixed garnet crystals. This task is not yet clarified for the application important Lu-based garnets.

In the present paper, the site occupations of Al and Ga ions have been investigated in Ce-doped $Lu_3Al_{5-x}Ga_xO_{12}$ mixed garnet crystals (x = 0, 1, 2, 3, 4, 5) by using the $^{27}Al$ and $^{71}Ga$ solid-state Nuclear magnetic resonance (NMR) technique. NMR, providing a unique approach to quantify the site occupancy in a material (in form of both liquid and solid state),[22] is an optimal tool to accomplish such a task. In $Lu_3Al_{5-x}Ga_xO_{12}$, both Al and Ga cations are suitable for NMR measurements as it was demonstrated for the $Y_3Al_{5-x}Ga_xO_{12}$ solid solutions.[19,20,23] Essential advantage of NMR over other methods sensitive to ionic occupation in lattice, for instance, x-ray diffraction, is its universality, insensitivity to long-range atomic ordering and the ability to provide information concerning local deformations arising due to slightly different ion sizes. To the best of our knowledge, the x-ray diffraction determination of the Al and Ga distribution over the cation sites was successfully achieved only for the high quality single crystals of $Y_3Al_{5-x}Ga_xO_{12}$ solid solutions.[17,18]

Our experimental study is accompanied by theoretical calculations based on the density functional theory (DFT) in order to investigate the tendency of the Al and Ga distribution over different positions in garnet structure and to explain unusual variation of the electric field gradients (EFG) in the $Lu_3Al_{5-x}Ga_xO_{12}$ lattice which do not obey simple point charge model.

## 2. Methods

The crystals were grown by the micro-pulling-down method with radiofrequency inductive heating.[24] An iridium crucible with a die of 3 mm in diameter was used. The growth was performed under $N_2$ atmosphere using <111> and <100> oriented $Y_3Al_5O_{12}$ single crystal as a seed. The starting materials were prepared by mixing 4N purity $Lu_2O_3$, $Al_2O_3$ and $Ga_2O_3$ powders. The Ce was added in the concentration 0.1 at.%. The crystals were in the form of rods with the diameter of 3 mm and length of 2−3 cm.

$^{27}$Al and $^{71}$Ga static NMR spectra were measured at room temperature using Bruker "Avance II" spectrometer at the Larmor frequency $\nu_L$ = 104.28 and 122.06 MHz, respectively. Magic-angle spinning (MAS) spectra of the both nuclei were measured using "Avance III HD" spectrometer equipped with high speed MAS probe at the Larmor frequency $\nu_L$ = 130.40 and 152.62 MHz for $^{27}$Al and $^{71}$Ga, respectively. The single pulse sequence with RF field strength of 110 KHz and short pulses of 0.38 µs, corresponding to the rotation of the net magnetization by the angle π/12, was applied to ensure the correctness of quantitative measurements of $^{27}$Al MAS NMR spectra corresponding to aluminum location at lattice positions with different values of the quadrupole coupling constant. From 512 up to 3072 scans with the recycle delay of 80 s were accumulated for each sample depending on the Al concentration. Spectral width of 312 KHz with spinning speed of 22 kHz was used to observe an undistorted central transition of $^{27}$Al in both the tetrahedral and octahedral positions. $^{27}$Al NMR spectra are referenced to external standard $Al(NO_3)_3$.

$^{71}$Ga static NMR spectra were measured utilizing the 90x−τ−90y−τ spin echo pulse sequence. Four phase (*xx*, *xy*, *x-x*, *x-y*) 'exorcycle' phase sequence was used to form echoes with minimal distortions due to anti-echoes, ill-refocused signals and piezo-resonances.[25] In the case of $^{71}$Ga NMR spectrum (linewidth up to 300 kHz for the 1/2 ↔ -1/2 central transition), only selective excitation takes place, but to exclude the influence of the angle dependence of quadrupolar interactions on the signal intensity, π/8 pulse lengths of 0.5 µs still were used with 5 s delays between scans. The $^{71}$Ga MAS NMR spectra were measured at rotation speed of 20 KHz by single pulse sequence with the pulse length of 0.8 µs and recycle delay of 15 s.

The electronic structures of studied garnets were calculated within the density functional theory by means of full-potential all-electrons augmented plane wave + local orbitals method as implemented in WIEN2k code.[26] The radii of atomic spheres in atomic units for Lu, Al, Ga, and O were 2.35, 1.70, 1.75, and 1.61, respectively. The atomic coordinates as well as the lattice parameters were optimized for each structure with respect to the total energy and within the set space group symmetry. A special care was taken to check how the geometry optimization (lattice parameters and the atomic forces) and also the key calculated quantities (electric field gradients and density of states) converge with the size of the basis set and with the number of k-points. The final appropriate value for the $RK_{max}$ parameter was 7.0, determining the matrix size about 9000 (double the size for the structures without I-centration). The number of k-points in the irreducible part of Brillouin zone was between 14 and 45, providing mesh 6×6×6 for each structure. The charge density and potentials were Fourier expanded up to largest k-vector $G_{max} = 14$ $Ry^{1/2}$. As the approximation to exchange-correlation potential, PBE variant[27] of generalized gradient approximation (GGA) was used.

## 3. Results and discussion

Al and Ga ions can occupy two different crystallographic positions in the garnet structure: tetrahedral (Al(IV), Ga(IV)) and octahedral (Al(VI), Ga(VI)) oxygen environment.[16] In the $Lu_3Al_{5-x}Ga_xO_{12}$ solid solution, both these cations are suitable for NMR measurements. The most easily measurable is $^{27}Al$ isotope, which has nuclear spin $I = 5/2$ and natural abundance 100%. Ga has two isotopes $^{69}Ga$ and $^{71}Ga$ with the spin 3/2 and natural abundance 60% and 40%, respectively. All these nuclei possess a quadrupole moment $eQ$, which interacts with the electric field gradient (EFG) generated by surrounding ions. The strength of the quadrupole interaction of the nuclei is usually expressed in terms of the quadrupole coupling constant $C_q = e^2 Q V_{zz} / h$, and the asymmetry parameter of the EFG tensor $\eta = \dfrac{V_{xx} - V_{yy}}{V_{zz}}$, where $e$ is the electron charge and $V_{ii}$ are the components of EFG tensor in its principal axes system (PAS) with $V_{zz}$ the largest component. In the $Ia\overline{3}d$ garnet structure,[16] there are 24 structurally equivalent tetrahedral and 16 structurally equivalent octahedral positions in the unit cell available for Al/Ga cations. At both types of sites EFG tensor has axial symmetry ($\eta = 0$) and the main axis of the EFG tensor is aligned along the <100> and <111> cubic directions for the tetrahedral and octahedral sites, respectively.[28,29] In the case of solid solution, the axial symmetry of the EFG tensor is obviously broken, as part of Al ions is substituted for Ga ions.

Quantitatively, all measured spectra of the static samples were interpreted by using the following spin Hamiltonian:[30,31]

$$\nu_m^{(1)} = \nu_L + (m - \frac{1}{2})\nu_Q \left\{ \left[ \left( 3\cos^2\theta - 1 \right)/2 \right] + \frac{1}{2}\eta \sin^2\theta \cos 2\varphi \right\} \qquad (1)$$

$$\nu_{1/2}^{(2)} = \nu_L - \frac{\nu_Q^2}{16\nu_L}\left( I(I+1) - \frac{3}{4} \right) f_\eta(\theta,\varphi) + \nu_L \left[ 1 + \delta_{iso} + \delta_{ax}\left( 3\cos^2\theta - 1 \right) + \delta_{aniso} \sin^2\theta \cos 2\varphi \right], \quad (2)$$

where $\nu_Q = \dfrac{3C_q}{2I(2I-1)}$ is the quadrupole frequency. $\theta$, $\varphi$ are the Euler angles of external magnetic field as referred to the principal x,y,z axis system of the EFG tensor. Eq. (1) describes the frequency locations of satellite transitions as first-order perturbation to the Larmor frequency $\nu_L$. Consequently, frequency shift related to magnetic-field screening at nucleus (chemical shift) is small compared to the quadrupole frequency and is neglected. On the other hand, shift of the central transition is only a second-order effect on the quadrupole frequency (Eq. (2)). Therefore, here the chemical shift mechanism is essential. Its contribution to the central transition frequency is described by the second term in Eq. (2), where $\delta_{iso}$, $\delta_{ax}$ and $\delta_{aniso}$ are isotropic, axial and anisotropic components of the chemical shift (CS) tensor, principal axis system of which is assumed to coincide with coordinate system of EFG tensor. The function $f_\eta(\theta,\varphi)$ has a cumbersome form and its actual expression is presented in Ref. 30. Interpretation of the MAS spectra obtained under rotation conditions was performed using the time dependent spin Hamiltonian, which includes all the interactions mentioned above. Its full analytic representation can be found elsewhere.[31]

### 3.1. $^{27}$Al NMR

As an example, $^{27}$Al static NMR spectra of grinded $Lu_3Al_5O_{12}$ and $Lu_3Al_2Ga_3O_{12}$ single crystals are presented in Fig. 1. NMR spectrum of $Lu_3Al_5O_{12}$ has, besides of the (1/2 ↔ -1/2) high intensity signal belonging to the central transition (CT), well pronounced singularities corresponding to the (±3/2 ↔ ±1/2) satellite transitions (ST) of both Al(IV) and Al(VI), which can be used for calculation of the quadrupole parameters. On the other hand, in the mixed $Lu_3Al_{5-x}Ga_xO_{12}$ compounds the singularities from the satellite transitions are completely smeared out (see the bottom spectrum in Fig. 1) due to the distribution of quadrupole parameters in these samples: the local environment of Al varies from site to site because of the Ga substitutions. The linewidth of the CT for all samples is about 20 kHz.

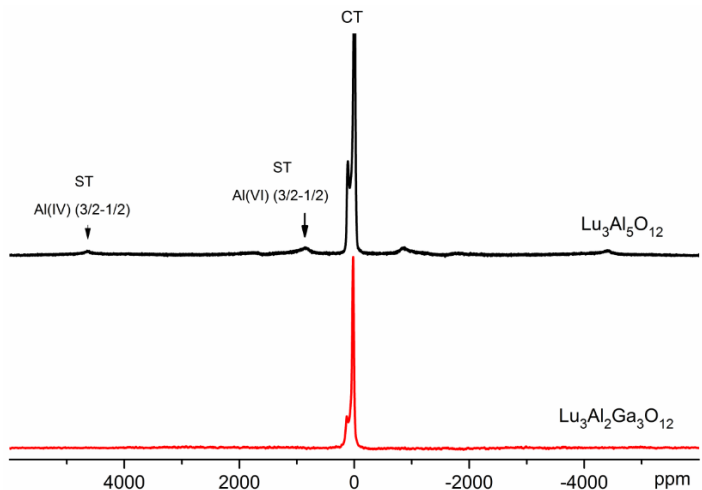

Fig. 1. $^{27}$Al static NMR spectrum of $Lu_3Al_5O_{12}$ and $Lu_3Al_2Ga_3O_{12}$. Arrows show the singularities corresponding to the ST's.

$^{27}$Al MAS NMR spectra of all samples, obtained with the spinning rate of 22 kHz, are presented in Fig. 2. This rotation frequency exceeds the linewidth of the CT of static spectrum, and so all magnetization from the CT is contained in two narrow lines corresponding to the CT of Al(IV) and Al(VI), which are separated due to the difference in chemical shifts.[28,29] A set of spinning side bands (ssb) originating from the satellite transitions is also present in MAS NMR spectra.

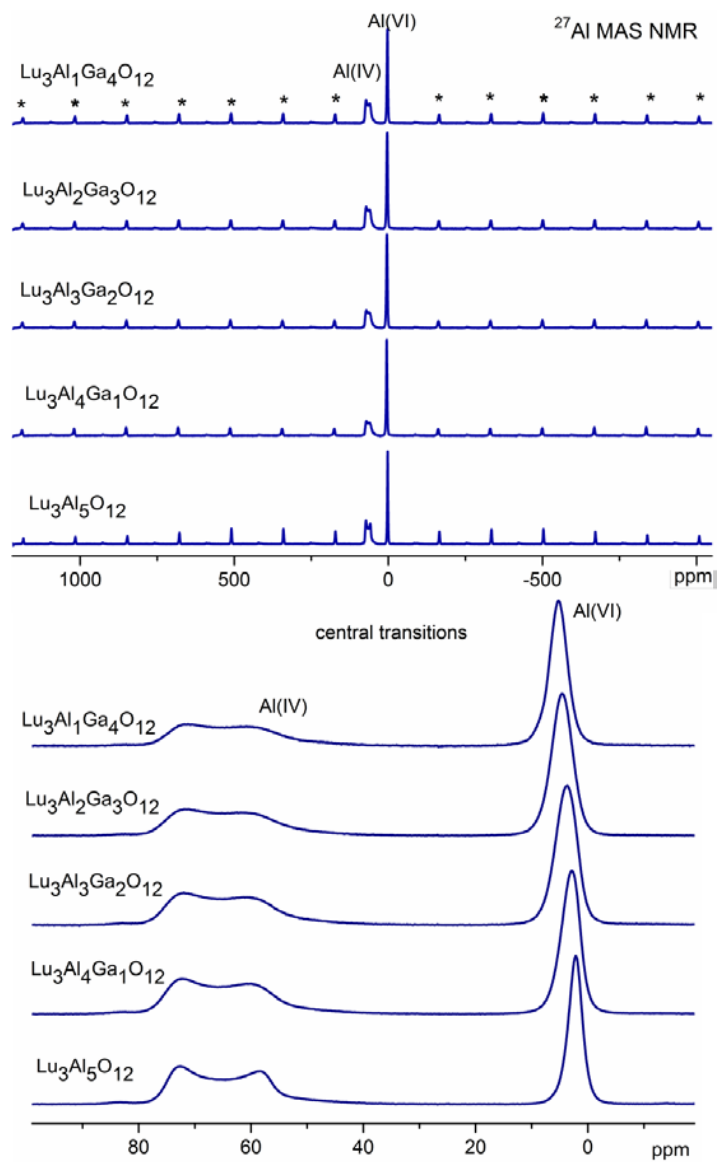

Fig. 2. Top panel: $^{27}$Al MAS NMR spectra of Lu$_3$Al$_{5-x}$Ga$_x$O$_{12}$ compounds obtained at rotation speed of 22 kHz, with the spinning sidebands denoted by the asterisks. Bottom panel: frequency region of the CT is zoomed in.

Spectral line corresponding to the Al(IV) CT has a pronounced shape due to strong quadrupole interactions (Fig. 2, panel (a)). Substitution of Al by Ga in the mixed compounds leads to the distribution of the EFG values on Al nuclei, which results in some smearing of the spectral line edges (Fig. 2, panel (b)). However, the spectral line corresponding to the central transition of Al(IV) still bears the information concerning the quadrupole interactions. Quadrupole interaction parameters obtained after fitting of the Al(IV) central transition line for all compounds are listed in Table 1. It can be seen that partial substitutions of Al by Ga do not change the mean EFG value, which evidences that Al oxygen tetrahedrons remain mainly unchanged.

Table 1. Al(IV) NMR parameters of the mixed $Lu_3Al_{5-x}Ga_xO_{12}$ compounds determined from MAS NMR spectra.

| Compound | $\delta_{iso}$ (ppm) | Cq (kHz) | η |
|---|---|---|---|
| $Lu_3Al_5O_{12}$ | 79.2 | 6330 | 0 |
| $LuAl_4Ga_1O_{12}$ | 79.2 | 6270 | 0.02 |
| $LuAl_3Ga_2O_{12}$ | 78.8 | 6200 | 0.03 |
| $LuAl_2Ga_3O_{12}$ | 78.5 | 6200 | 0.004 |
| $LuAlGa_4O_{12}$ | 78.2 | 6230 | 0.01 |

Because the EFG values at octahedral sites are almost six times smaller than those at tetrahedral sites (their actual values will be determined below), CT of Al(VI) is completely averaged to one narrow line of Gaussian shape. Its full width at half maximum (FWHM) changes from 2.9 ppm at x = 0 up to 4.6 ppm at x = 2 (Table 2) indicating that maximum distribution of NMR parameters takes place for the $Lu_3Al_3Ga_2O_{12}$ solid solution. One can also see that the isotropic chemical shift, determined directly from the MAS spectra, gradually increases from 2.2 ppm in $Lu_3Al_5O_{12}$ up to 5.3 ppm in $Lu_3AlGa_4O_{12}$.

MAS NMR line of the Al(VI) central transition has no singularities typical for quadrupole interactions. Therefore, the Al(VI) quadrupole parameters were calculated through the full fit of the spinning side band (ssb) manifold originated from the satellite transitions (±3/2 ↔ ±1/2, ±5/2 ↔ ±3/2). In contrast to $Lu_3Al_5O_{12}$, the shape of ssb manifold in the spectra of mixed $Lu_3Al_{5-x}Ga_xO_{12}$ compounds is smeared out due to the distribution of EFG parameters. To increase the precision in the determination of the Al(VI) quadrupole parameters, a larger number of ssb was accumulated in the region -0.55 – 0.55 MHz with rotation speed of 5 KHz (Fig. 3). Intensities of the sidebands were also corrected for the Q-factor of the probe by a simple Lorentz function.

Taking into account that ssb envelope to some extent (but not completely) mimics the shape of the static spectra, some starting values of the quadrupole parameters were obtained by fitting the ssb envelope with the line corresponding to NMR spectrum of a static sample. As it was shown in Ref. 32 along with our measurement performed for $Lu_3Al_5O_{12}$, quadrupole parameters obtained from such fitting have lower values than the real ones. But, as the calculation of the static spectrum is fast, they can be used as good starting values for further fitting of the sidebands manifold, and give correct tendency of parameters behavior with increasing degree of cation substitution.

The spectrum of the satellite transitions (ST) of static samples was simulated by using spin Hamiltonian (1). Distribution of the quadrupole frequencies arising from EFG fluctuations was considered as a Gaussian function. The satellite transitions spectrum was obtained by subsequent integration over all mutual orientations of EFG tensor axis and magnetic field:

$$I(\nu) = \sum_m \int_0^\pi \int_0^{\pi/2} \sqrt{\frac{\ln 2}{\pi}} \frac{\sin\theta}{\Delta} \exp\left[-\left(\frac{\nu - \nu_m^{(1)}(\theta,\varphi)}{\Delta}\right)^2\right] d\theta d\varphi, \tag{3}$$

where the parameter $\Delta$ characterizes the width of the quadrupole frequency distribution and the frequency $\nu_m^{(1)}(\theta,\varphi)$ is defined by Eq. (1).

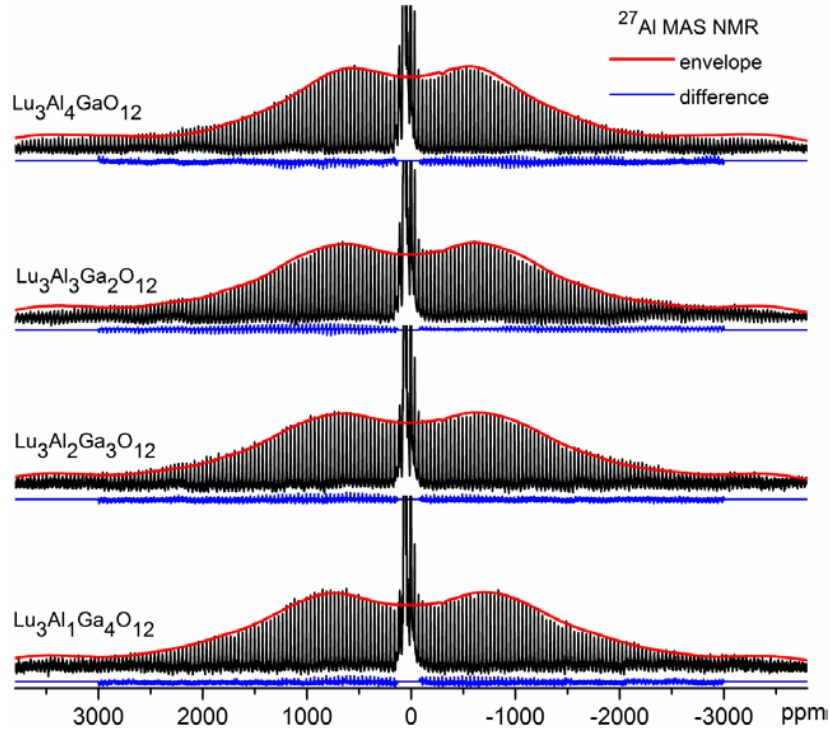

Fig. 3. $^{27}Al$ MAS NMR spectra of $Lu_3Al_{5-x}Ga_xO_{12}$ obtained at rotation speed of 5 kHz together with sidebands-envelope simulation (red lines) and the difference (blue lines) between the measured spectrum and the best fit of all sidebands manifold (only the region corresponding to Al(VI) ST sidebands is presented in the difference curve). Sidebands intensities were corrected taking into account the Q-factor of the probe.

Subsequent simulation of the Al(VI) ssb manifold was performed taking into account that actual EFG distribution originates from the random distribution of Al and Ga atoms located in the nearest environment of Al(VI). Therefore, a finite number of different EFG values instead of smooth EFG distribution can be used during the simulation of the spectra. Farther, fixed parameters for Al(IV) were used as well during the simulation. Mean values of the quadrupole parameters obtained in this way are listed in Table 2. To illustrate the simulation precision, the differences between the experimental spectra and simulated spectra are presented in Fig. 3 by blue solid lines for the region where the main signal from Al(VI) ST is located, farther, Al(IV) ST and the CT region were excluded from the calculation.

Table 2. Al(VI) NMR spectral parameters of the mixed $Lu_3Al_{5-x}Ga_xO_{12}$ compounds determined from MAS NMR spectra of the CT (A), simulation of the ssb envelope (B), and using the direct fit of all spinning side bands (C).

| Compound | $\delta_{iso}$ (ppm) | FWHM (ppm) | $C_q$ (kHz) | η | $\delta_{iso}$ (ppm) | $C_q$ (kHz) | η |
|---|---|---|---|---|---|---|---|
| | A | | B | | | C | |
| $Lu_3Al_5O_{12}$ | 2.2 | 2.9 | 1100 | 0 | 2.6 | 1230 | 0 |
| $Lu_3Al_4Ga_1O_{12}$ | 2.9 | 3.9 | 1400 | 0.3 | 3.8 | 1560 | 0.4 |
| $Lu_3Al_3Ga_2O_{12}$ | 3.8 | 4.6 | 1550 | 0.35 | 4.6 | 1690 | 0.4 |
| $Lu_3Al_2Ga_3O_{12}$ | 4.7 | 4.4 | 1650 | 0.35 | 5.4 | 1820 | 0.4 |
| $Lu_3AlGa_4O_{12}$ | 5.3 | 3.7 | 1750 | 0.3 | 6.3 | 1900 | 0.4 |

It can be seen that, in contrast to the Al(IV) tetrahedral sites (Table 1), the EFG parameters at Al(VI) octahedral sites essentially depend on Ga substitution, which indicates that oxygen octahedrons experience marked deformations with substitutions of Al by Ga. The degree of octahedron's deformations grows up steadily with the degree of Al → Ga substitution. An increase in the average values of the chemical shift with gradual substitution of Al by Ga is in good agreement with the electronegativities of Al and Ga atoms, which have the Pauling values of 1.61 and 1.81, respectively. Substitution of Al by more electronegative Ga atoms decreases the total charge on the oxygen framework and results in additional transfer of the electron density from Al to O atoms. This decreases the net charge on Al atoms, which results in lower shielding of the external magnetic field by Al electronic shell. Also, because Ga prefers to occupy the tetrahedral positions, Al ions in the mixed structure are predominantly surrounded by Ga ions leading to essential change of the EFG and chemical shift parameters as compared to LuAG compound. Thus, in the case of disordered $LuAl_3Ga_2O_{12}$, a larger number of possible Al environments results in a broader distribution of the chemical shifts.

Because [27]Al MAS NMR spectral lines from both Al structural positions are well separated by the chemical shift, we can directly calculate the occupation numbers of Al located in the different positions, which, in principle, should be equal to the ratio of integral intensities of Al(IV) and Al(VI) spectral lines. Such ratios calculated for CT's are presented in Table 3. It can be seen that the Al(IV)/Al(VI) intensities ratio obtained for pure $Lu_3Al_5O_{12}$ does not coincide with value 1.5 expected for the stoichiometric compound. This significant deviation is caused by the overlap of the central transition line with lines originating from satellite transitions, which have high enough intensity at this rotation speed (Fig. 2). Intensities of the ST lines are different for Al(IV) and Al(VI) sites due to the difference in their quadrupole parameters, which determines the shape of the side bands manifold and their contribution to the overlap with the CT.

In order to obtain correct values of the Al(IV)/Al(VI) ratio one should subtract part of the magnetization (spectral intensity) belonging to ST from CT. The correction of the integral intensities was done by fitting of the experimentally accumulated spectrum to obtain the quadrupolar parameters for both Al(IV) and Al(VI) sites, and consequently, simulating the entire spectrum in the range up to $4 \times v_Q$. This takes into account all the side bands from all the transitions not directly visible in the experimental spectrum, so the ratio of the integral intensities of the entire lines gives the correct Al occupation numbers.

Following this algorithm we firstly checked the Al occupation numbers of $Lu_3Al_5O_{12}$. Its MAS spectrum can be well fitted with the following parameters (in a good agreement with published values for this compound[33]): $C_q$ = 1230 and 6250 kHz for the Al(VI) and Al(IV), respectively and $\eta$ = 0 for both sites. Calculated Al(IV)/Al(VI) ratio for $Lu_3Al_5O_{12}$ now gives 1.51(1), which within the error range coincides with the expected value 1.5.

Quadrupole parameters obtained for mixed compounds allowed us to calculate the actual Al(IV)/Al(VI) ratio from the MAS NMR spectra. Al(IV)/Al(VI) ratios obtained after simulation of the spectra are listed in Table 3.

Table 3. Occupation numbers (Al(IV)/Al(VI) ratio) of the mixed $Lu_3Al_{5-x}Ga_xO_{12}$ compounds determined directly from $^{27}Al$ MAS spectra by the intensity ratio of CT (A), and obtained after the simulation of full spectra (B).

| Compound | Al(IV)/Al(VI) ratio | |
|---|---|---|
| | A | B |
| $Lu_3Al_5O_{12}$ | 1.34 | 1.51 |
| $Lu_3Al_4GaO_{12}$ | 1.08 | 1.19 |
| $Lu_3Al_3Ga_2O_{12}$ | 0.92 | 1.01 |
| $Lu_3Al_2Ga_3O_{12}$ | 0.78 | 0.84 |
| $Lu_3AlGa_4O_{12}$ | 0.72 | 0.76 |

We note that, taking into account very smooth shape of the sidebands envelope in the case of mixed $Lu_3Al_{5-x}Ga_xO_{12}$ compounds, almost the same correct Al(IV)/Al(VI) ratio can be obtained from MAS spectra just by subtraction of the intensity of first sideband (obtained at rotation of 22 KHz) from the intensity of the central line.

## 3.2. $^{71}$Ga NMR

$^{71}$Ga static NMR spectrum of $Lu_3Ga_5O_{12}$ sample obtained from the grinded single crystal is presented in Fig. 4 together with its best fit. The spectrum was analyzed similarly as in.[23] Central transition spectral line of Ga(VI) locates approximately between 200 and -200 ppm and was fitted using $C_q$ = 5300 kHz and $\eta$ = 0. Central transition spectrum of Ga(IV) spreads from 800 to -800 ppm and yields $C_q$ = 13650 kHz and $\eta$ = 0.

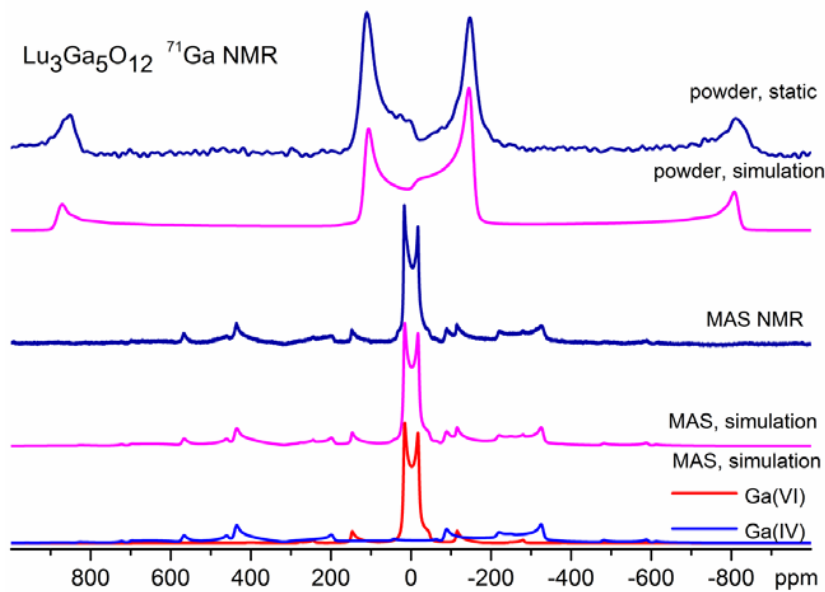

Fig. 4. $^{71}$Ga NMR static and MAS spectra (rotation speed 20 kHz) of grinded $Lu_3Ga_5O_{12}$ single crystal (dark lines), and their best fit (color lines).

$^{71}$Ga MAS NMR spectrum of $Lu_3Ga_5O_{12}$ is also presented in Fig. 4. Spectral line of the Ga(VI) central transition can be clearly seen near 0 ppm. At this rotation speed, its lineshape has distinct singularities related to quadrupole interaction, which allows calculation of the quadrupole parameters of Ga(VI). Ga(IV) central transition cannot be well resolved at available speeds of rotations as it spreads over 200 kHz for the static spectrum. On the other hand, all the features presented in the $^{71}$Ga MAS NMR spectrum can be unambiguously fitted with good accuracy even for the spectrum obtained with the rotation 20 kHz (Fig. 4). This is rather in contrast with the fitting of the static spectrum, where different combinations of quadrupole interactions and chemical shift anisotropies can result in similar spectra. We note that quadrupole

parameters obtained after simulation of the MAS spectra lead to an almost correct lineshape of the static spectrum, which validates the analysis of MAS spectra for all samples.

$^{71}$Ga MAS NMR spectra of all studied compounds are presented in Fig. 5. Quadrupole parameters obtained from the best fit of these spectra are summarized in Table 4.

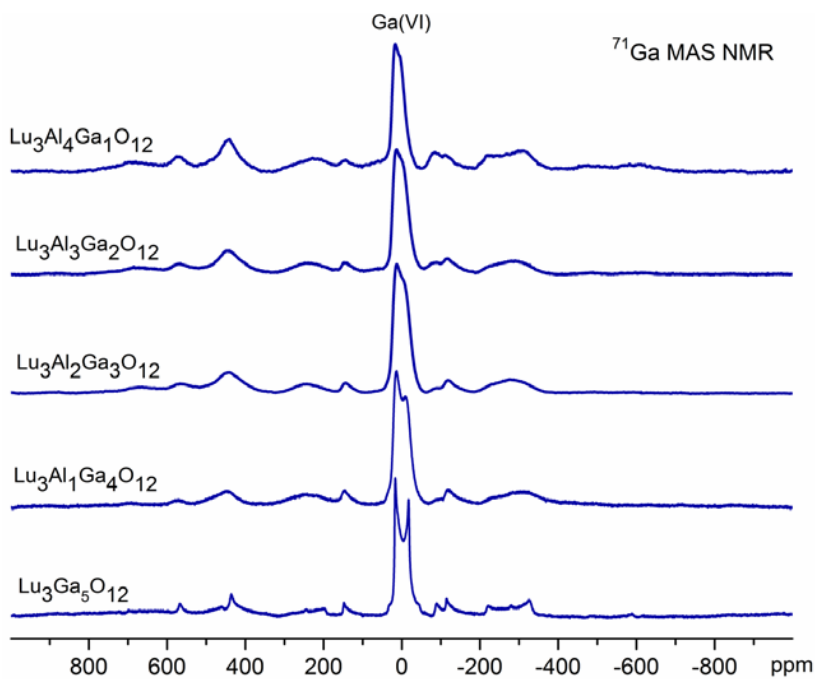

Fig. 5. $^{71}$Ga MAS NMR spectra of $Lu_3Al_{5-x}Ga_xO_{12}$ obtained at rotation speed of 20 kHz.

Table 4. $^{71}$Ga NMR quadrupole parameters of the mixed $Lu_3Al_{5-x}Ga_xO_{12}$ compound.

|  | Ga(IV) | | | Ga(VI) | | |
|---|---|---|---|---|---|---|
| Compound | $C_q$ (kHz) | $\eta$ | $\delta_{iso}$ (ppm) | $C_q$ (kHz) | $\eta$ | $\delta_{iso}$ (ppm) |
| $Lu_3Al_4GaO_{12}$ | 13500 | 0 | 250 | 4450 | 0 | 29 |
| $Lu_3Al_3Ga_2O_{12}$ | 13700 | 0 | 250 | 4950 | 0 | 29 |
| $Lu_3Al_2Ga_3O_{12}$ | 13800 | 0 | 260 | 5050 | 0 | 30 |
| $Lu_3AlGa_4O_{12}$ | 13900 | 0 | 270 | 5330 | 0 | 30 |
| $Lu_3Ga_5O_{12}$ | 13700 | 0 | 250 | 5310 | 0 | 28 |

With increasing content of gallium the quadrupole constant $C_q$ for Ga located in octahedral environment increases, while the quadrupole parameters of Ga in the tetrahedral environment are almost unchanged. It should be noted that very large differences between $C_q$ values of Ga and Al atoms located in similar environments are connected with different values of Sternheimer antishielding factors ("a measure of the magnification of eq due to distortion of the inner electrons close to a nucleus"[22]) for these atoms, equal -3.6 and -17 for Al and Ga, respectively.

In order to obtain the Ga(IV)/Ga(VI) occupation ratio and to take into account possible base-line distortions of the very broad Ga(IV) lines in the $^{71}$Ga MAS spectra, $^{71}$Ga static NMR spectra of $Lu_3Al_{5-x}Ga_xO_{12}$ single crystals were measured (Fig. 6).

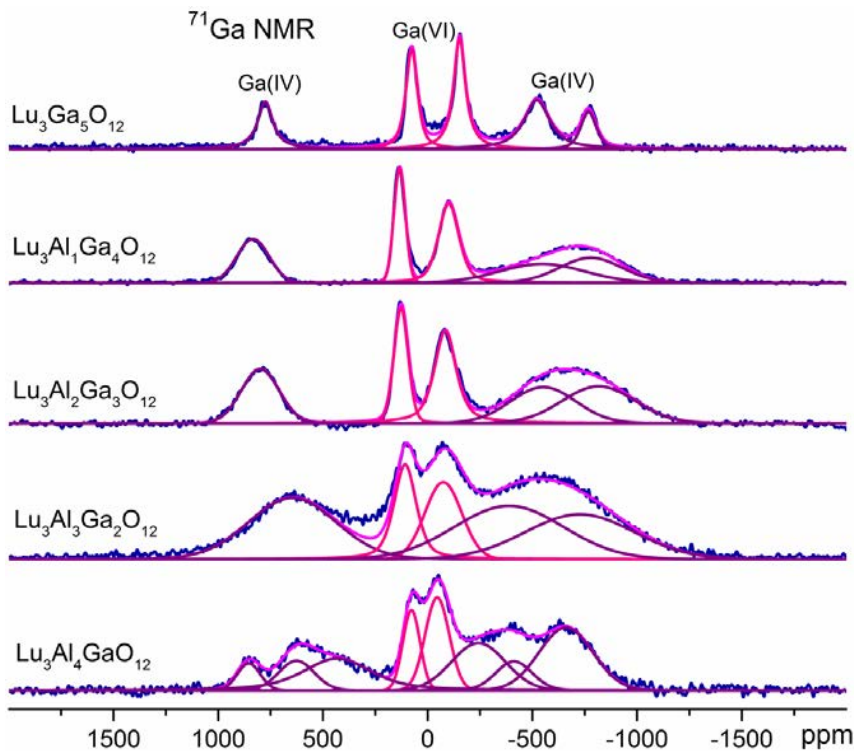

Fig. 6. $^{71}$Ga single crystal NMR spectra of the $Lu_3Al_{5-x}Ga_xO_{12}$ compounds. Solid lines of different colors are the Gaussian decomposition of the spectra into separate contributions from the Ga(VI) and Ga(IV) sites.

Assignment of the spectral lines originating from the Ga(IV) and Ga(VI) sites was made on the basis of the data obtained from MAS spectra while taking into account the essential difference between Ga(IV) and Ga(VI) quadrupole parameters (Table 5). Two groups of lines can be distinguished in the spectra of the single crystals: a group of lines shifting under crystal rotation in the range from approximately -200 to 200 ppm was assigned to Ga(VI), while another group of lines shifting with crystal rotation in the range from approximately -1000 to 1000 ppm was assigned to Ga(IV) due to much larger quadrupole constant.

By choosing an appropriate crystal orientation, $^{71}$Ga NMR lines of Ga(IV) can be well resolved from the Ga(VI) sites even for broadened spectral lines. Such spectra of all the $Lu_3Al_{5-x}Ga_xO_{12}$ mixed compounds (x = 1−5) are shown in Fig. 6. Ga(IV)/Ga(VI) occupation numbers calculated from the single-crystal spectra are listed in Table 5, together with the Ga(IV)/Ga(VI) ratio recalculated from Al(IV)/Al(VI) ratio (obtained earlier in section 3.1) on the basis of the following relations:

$$(N_{Al(IV)} + N_{Ga(IV)})/(N_{Al(VI)} + N_{Ga(VI)}) = 3/2 \qquad (4)$$

$$(N_{Al(IV)} + N_{Al(VI)})/(N_{Ga(IV)} + N_{Ga(VI)}) = (5-x)/x,$$

where $N_{Al(IV)}$, $N_{Al(VI)}$ and $N_{Ga(IV)}$, $N_{Ga(VI)}$ are the numbers of Al and Ga atoms in the tetrahedral and octahedral positions, respectively.

Table 5. Ga(IV)/Ga(VI) occupation ratio determined from $^{71}$Ga and $^{27}$Al NMR spectra in the mixed $Lu_3Al_{5-x}Ga_xO_{12}$ compounds.

| Compound | Ga(IV)/Ga(VI) | |
|---|---|---|
| | Determined from $^{71}$Ga NMR | Recalculated from $^{27}$Al MAS NMR |
| $Lu_3Al_5O_{12}$ | - | - |
| $Lu_3Al_4Ga_1O_{12}$ | 3.6 | 4.8 |
| $Lu_3Al_3Ga_2O_{12}$ | 3.01 | 2.9 |
| $Lu_3Al_2Ga_3O_{12}$ | 2.15 | 2.29 |
| $Lu_3AlGa_4O_{12}$ | 1.82 | 1.8 |
| $Lu_3Ga_5O_{12}$ | 1.49 | - |

We note, however, that due to the broad spectral lines of the $^{71}$Ga NMR spectra, the accuracy of the Ga(IV)/Ga(VI) ratio determination using the $^{71}$Ga NMR spectra is much lower than that recalculated from the $^{27}$Al MAS NMR, and can thus be regarded as a supplementary information only. Nevertheless, some additional information can be extracted from the single-crystal spectra. Considering the garnet structure with allowed $Ia\bar{3}d$ symmetry for all the single crystals, the main component of the EFG tensor of Ga(IV) should be oriented along the $\bar{4}$ axis, which is aligned along one of the simple [100], [010] and [001] cubic directions. This results in three magnetically inequivalent Ga(IV) sites in a single crystal, with $\eta = 0$ due to the presence of the improper rotation axis, and so only three lines should be present in the spectrum of Ga(IV). The presence of at least four distinct lines belonging to Ga(IV) on the spectrum of $LuAl_4Ga_1O_{12}$ in Fig. 6, with $\eta = 0$ (obtained from MAS) evidences that partial substitution of Al by Ga changes the orientation of the local axis for some tetrahedral sites, leaving them otherwise undistorted. These data together with the data obtained for Al allow us to suppose that partial substitution of Al by Ga results in rotation of oxygen tetrahedrons with simultaneous deformation of oxygen octahedrons.

To characterize more precisely the Al and Ga ions occupations, we have calculated fractional occupation numbers, defined as

$$f_{Al}^{tet} = Al(IV)/(Al(IV) + Al(VI)) \text{ and}$$

$$f_{Al}^{oct} = Al(IV)/(Al(IV)+Al(VI))$$

for Al ions and similarly for Ga ions (Fig. 7). One can see in particular that at x = 1, only 20% of the Ga ions occupy the octahedral sites whereas 80% occupy the tetrahedral sites. Both $f_{Ga}^{tet}$ and $f_{Ga}^{oct}$ linearly tend to their normal values of 60 and 40% as the Ga concentration increases. Similar result was obtained in $Y_3Ga_xAl_{5-x}O_{12}$ by both x-ray diffraction[17] and NMR[20] methods supporting that Al and Ga site occupation in garnets, obviously, does not depend on rare earth dodecahedral cations.

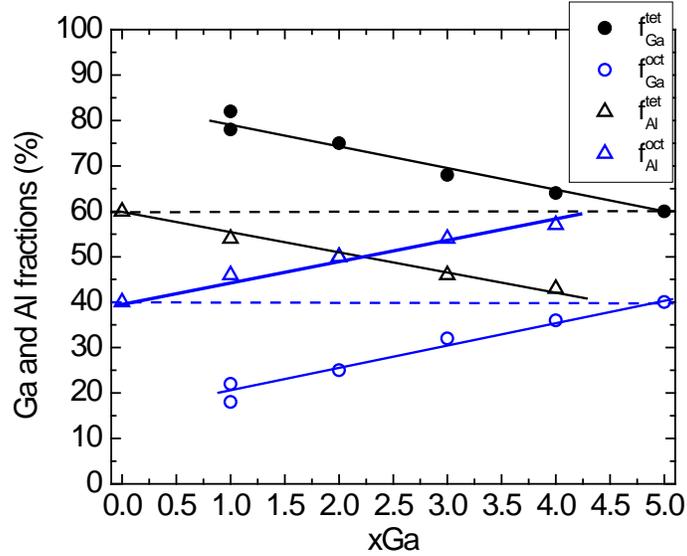

Fig. 7. Fractional occupation parameters of Ga and Al ions in the $Lu_3Ga_xAl_{5-x}O_{12}$ mixed crystals as a function of the total Ga content. Dashed lines correspond to random distribution of Al and Ga over tetrahedral and octahedral sites: $f_{Al}^{tet} = f_{Ga}^{tet} = 60\%$ and $f_{Al}^{oct} = f_{Ga}^{oct} = 40\%$.

### 3.3. DFT calculations

Electronic structures of measured compounds with varying gallium content *x* were modelled using single garnet unit cells $Lu_{24}Ga_{8x}Al_{40-8x}O_{96}$ containing 8 formula units. The number of possible arrangements of Al and Ga within one garnet unit cell increases rapidly when shifting away from the pure LuAG or LuGG composition (x=0 or 5). Therefore, besides designing the different distributions of Ga and Al cations in tetra- and octahedral sites, we also aimed to keep the space group symmetry reasonably high in order to make the calculations feasible. Along with the two pure garnet structures $Lu_3Al_5O_{12}$ and $Lu_3Ga_5O_{12}$, we calculated mixed compounds with content of Ga x=1, 2, 3, and 4, where for each *x* a small number of the most symmetric configurations were considered. All calculated structures together with their structural data are listed in Table 6.

Table 6. List of the structures considered in DFT calculations.

| Compound | Cations distribution over the sites: () = tetrahedral, [] = octahedral | Al(IV)/Al(VI) ratio | Space group | Lattice constant (a.u.) |
|---|---|---|---|---|
| $Lu_3Al_5O_{12}$ | $(Al_{24})[Al_{16}]$ | 1.5 | Ia-3d | a=22.72618 |
| $Lu_3Al_4GaO_{12}$ | $(Al_{16}Ga_8)[Al_{16}]$ | 1.0 | $I4_1/acd$ | a=22.83831 |
| | | | P-4 | a=22.83831[1) |
| | $(Al_{24})[Al_8Ga_8]$ | 3.0 | Ia-3 | a=22.86266 |
| | | | R32 | a=32.33268[1) c=39.59928 γ=120° |
| $Lu_3Al_3Ga_2O_{12}$ | $(Al_8Ga_{16})[Al_{16}]$ | 0.5 | $I4_1/acd$ | a=22.93970 |
| | $(Al_{16}Ga_8)[Al_8Ga_8]$ | 2.0 | Ibca | a=22.96419 |
| | $(Al_{24})[Ga_{16}]$ | ∞ | Ia-3d | a=22.98407 |
| $Lu_3Al_2Ga_3O_{12}$ | $(Al_{16}Ga_8)[Ga_{16}]$ | ∞ | $I4_1/acd$ | a=23.08864 |
| | $(Al_8Ga_{16})[Al_8Ga_8]$ | 1.0 | Ibca | a=23.07941 |
| | $(Ga_{24})[Al_{16}]$ | 0 | Ia-3d | a=23.04709 |
| $Lu_3AlGa_4O_{12}$ | $(Al_8Ga_{16})[Ga_{16}]$ | ∞ | $I4_1/acd$ | a=23.19772 |
| | | | P-4 | a=23.19772[1) |
| | $(Ga_{24})[Al_8Ga_8]$ | 0 | Ia-3 | a=23.16989 |
| | | | R32 | a=32.76718[1) c=40.13143 γ=120° |
| $Lu_3Ga_5O_{12}$ | $(Ga_{24})[Ga_{16}]$ | - | Ia-3d | a=23.302087 |

[1) For structures with lower symmetry the lattice parameters were not fully optimized but taken from the corresponding structure with higher symmetry

The quadrupole parameters, such as the parameters of electric field gradient tensor, $V_{zz}$ and asymmetry parameter η, computed for all these compounds are presented in Table 7 (where $V_{zz}$ is converted to $C_q$). EFG parameters are usually in a good agreement with experiment[34-36] and can thus be directly compared with the NMR results. It can be seen from Table 7 that preferable allocation of Ga either in octa- or tetrahedral position substantially influences the computed values of Ga(VI) $C_q$ (see e.g. data for $Lu_3Al_3Ga_2O_{12}$ compound, where $C_q$ varies from 3984 to 2622 kHz depending on the Al(IV)/Al(VI) ratio). For the calculated structures with the Al/Ga arrangement, chosen as close as possible to the real distribution defined by the NMR, the calculated $C_q$ values slightly underestimate the experimental ones but overall match the NMR experiment well and confirm our interpretation of the measured data (see Fig. 8).

Table 7. Calculated EFG parameters $C_Q$ (in kHz) and $\eta$ for Ga and Al nuclei in all considered structures.

| Compound | Cation distribution over () = tetrahedral, [] = octahedral sites | Al(IV)/Al(VI) | Al(IV) | | Al(VI) | | Ga(IV) | | Ga(VI) | |
|---|---|---|---|---|---|---|---|---|---|---|
| | | | $C_q$ | $\eta$ | $C_Q$ | $\eta$ | $C_Q$ | $\eta$ | $C_q$ | $\eta$ |
| $Lu_3Al_5O_{12}$ | $(Al_{24})[Al_{16}]$ | 1.5 | 5993 | 0 | 1045 | 0 | - | - | - | - |
| $Lu_3Al_4GaO_{12}$ | $(Al_{16}Ga_8)[Al_{16}]$ | 1.0 | 5766 | 0.1 | 1662 | 0.49 | 12568 | 0 | - | - |
| $Lu_3Al_4GaO_{12}$ | $(Al_{24})[Al_8Ga_8]$ | 3.0 | 6183 | 0.46 | 907 | 0 | - | - | 2976 | 0 |
| $Lu_3Al_3Ga_2O_{12}$ | $(Al_8Ga_{16})[Al_{16}]$ | 0.5 | 5608 | 0 | 1966 | 0.43 | 12195 | 0.1 | - | - |
| $Lu_3Al_3Ga_2O_{12}$ | $(Al_{16}Ga_8)[Al_8Ga_8]$ | 2.0 | 6040 | 0.58 | 1555 | 0.59 | 12897 | 0.37 | 3984 | 0.31 |
| $Lu_3Al_3Ga_2O_{12}$ | $(Al_{24})[Ga_{16}]$ | - | 6326 | 0 | - | - | - | - | 2622 | 0 |
| $Lu_3Al_2Ga_3O_{12}$ | $(Al_{16}Ga_8)[Ga_{16}]$ | - | 6182 | 0.11 | - | - | 13022 | 0 | 3847 | 0.38 |
| $Lu_3Al_2Ga_3O_{12}$ | $(Al_8Ga_{16})[Al_8Ga_8]$ | 1.0 | 5772 | 0.5 | 1785 | 0.47 | 12504 | 0.47 | 4474 | 0.32 |
| $Lu_3Al_2Ga_3O_{12}$ | $(Ga_{24})[Al_{16}]$ | - | - | - | 2031 | 0 | 11803 | 0 | - | - |
| $Lu_3AlGa_4O_{12}$ | $(Al_8Ga_{16})[Ga_{16}]$ | - | 5991 | 0 | - | - | 12686 | 0.09 | 4228 | 0.43 |
| $Lu_3AlGa_4O_{12}$ | $(Ga_{24})[Al_8Ga_8]$ | 0 | - | - | 1797 | 0 | 12138 | 0.41 | 4692 | 0 |
| $Lu_3Ga_5O_{12}$ | $(Ga_{24})[Ga_{16}]$ | - | - | - | - | - | 11514 | 0 | 4129 | 0 |

According to our calculations and the experiment, partial substitution of Al by Ga results in increase of $C_q$ values of both Al(VI) and Ga(VI), which indicates the distortion of oxygen octahedrons. Note, that the distortion of the octahedrons is accompanied by an increase of the lattice constants, which can be evidenced from both: our calculations performed for mixed $Lu_3Al_{5-x}Ga_xO_{12}$ compounds (Table 6), and x-ray diffraction data measured for $Y_3Al_{5-x}Ga_xO_{12}$ mixed compounds.[17,18] Distortion of oxygen octahedrons with increasing Ga content, in some cases, can be explained by non-uniform octahedron's environment arising after partial substitution of Al by Ga. However, in the case of $Lu_3Ga_5O_{12}$, $C_q$ for Ga(VI) has noticeably higher value as compared to $C_q$ for Ga(VI) in $Lu_3Al_4GaO_{12}$ (Tables 4 and 7), while $Lu_3Ga_5O_{12}$ has larger octahedral volume and, unlike $Lu_3Al_4GaO_{12}$, it has an absolutely uniform environment of Ga octahedrons. So, partial substitution of Al by Ga changes the symmetry of Ga(VI) local environment, resulting in increase of the octahedrons deformation with increase of the unit cell volume.

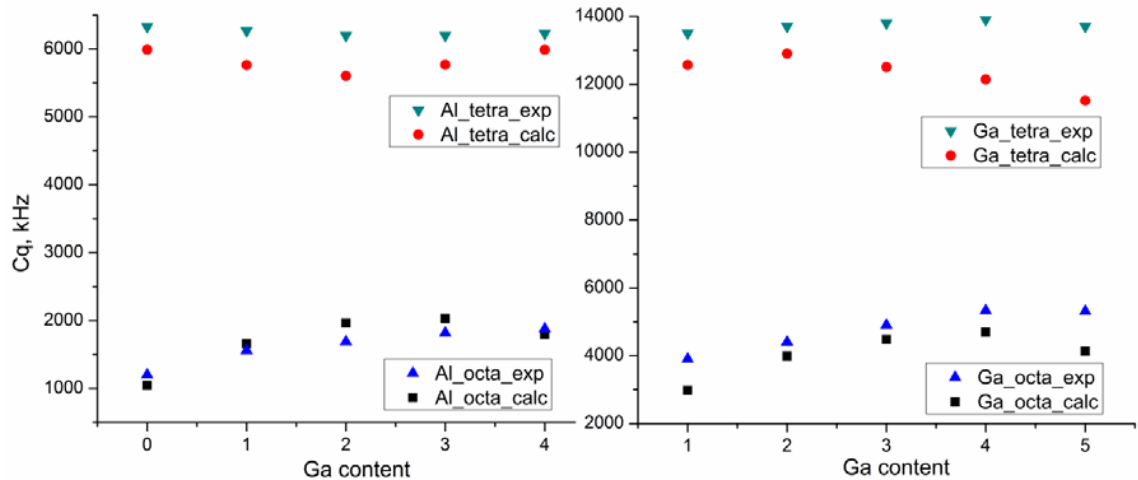

Fig. 8. Calculated quadrupole constants $C_q$ obtained from NMR experiment and DFT calculations in dependence on gallium content *x*.

Different distribution of Ga over tetrahedral and octahedral sites affects also the lattice constant (Table 6), which then slightly deviates from the values predicted by the Vegard's law (Fig. 9). In particular, the lattice constants of the configurations with the tetrahedral sites preferably occupied by Ga ions have somewhat lower values than those expected on the basis of the Vegard's law. Such deviation was experimentally found for $Y_3Al_{5-x}Ga_xO_{12}$ compounds[17] where on the basis of X-ray diffraction data it was shown that Ga preferably occupies the tetrahedral sites. On the other hand, the lattice constants of model structures, where Ga was forced to occupy the octahedral sites, display quite an opposite deviation from the Vegard's law (Fig. 8). Qualitatively, the same deviation from the Vegard's law was theoretically found in,[21] but just for two structures $Y_3Al_{5-x}Ga_xO_{12}$, with x = 0.6, where Ga occupied octahedral sites only, and x = 0.4, with Ga occupying tetrahedral sites, only.

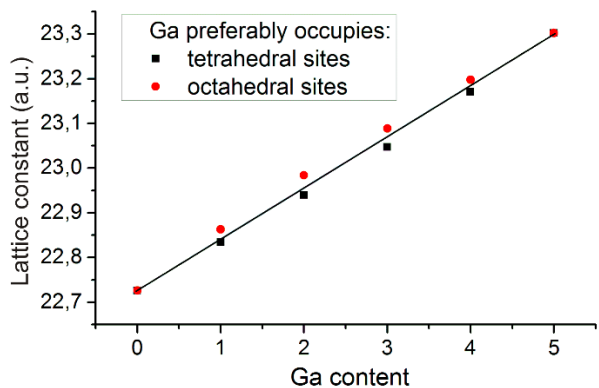

Fig. 9. Dependence of the lattice constant on Ga content. The solid line shows the lattice constants expected from the Vegard's law.

The deviation from the Vegard's law implies changes in cation-cation distances for different Ga distributions and influences the repulsive forces between them.[17] We have calculated the total energies for all studied configurations under condition that their unit cell parameters and atomic coordinates have been properly optimized, with remaining atomic forces well below 2 mRy/a.u. (Table 8). All the energies presented in Table represent the difference between the lowest energy for compound with some cations distribution and the energy for compound with specific arrangements of Al and Ga atoms over tetra- and octahedral sites; each difference is normalized by the number of Ga-Al pairs which should be swapped in order to obtain such particular arrangement of Al and Ga cations.

Table 8. The energy gain for swapping Ga-Al pair between the octahedral and tetrahedral positions for $Lu_3Al_{5-x}Ga_xO_{12}$ structures.

| Compound | Cations distribution over the sites: <br> () = tetrahedral, <br> [] = octahedral | Energy difference per Ga-Al pair (meV) |
|---|---|---|
| $Lu_3Al_5O_{12}$ | $(Al_{24})[Al_{16}]$ | - |
| $Lu_3Al_4GaO_{12}$ | $(Al_{16}Ga_8)[Al_{16}]$ | 0 |
| | | 333 |
| | $(Al_{24})[Al_8Ga_8]$ | 212 |
| | | 535 |
| $Lu_3Al_3Ga_2O_{12}$ | $(Al_8Ga_{16})[Al_{16}]$ | 0 |
| | $(Al_{16}Ga_8)[Al_8Ga_8]$ | 200 |
| | $(Al_{24})[Ga_{16}]$ | 195 |
| $Lu_3Al_2Ga_3O_{12}$ | $(Al_{16}Ga_8)[Ga_{16}]$ | 183 |
| | $(Al_8Ga_{16})[Al_8Ga_8]$ | 185 |
| | $(Ga_{24})[Al_{16}]$ | 0 |
| $Lu_3AlGa_4O_{12}$ | $(Al_8Ga_{16})[Ga_{16}]$ | 179 |
| | | 178 |
| | $(Ga_{24})[Al_8Ga_8]$ | 9 |
| | | 0 |
| $Lu_3Ga_5O_{12}$ | $(Ga_{24})[Ga_{16}]$ | - |

Our calculations clearly show, that for any given composition of the garnet (defined by gallium content $x$) the scenario Ga(IV)+Al(VI) is more favorable than the opposite Ga(VI)+Al(IV). The energy difference is about 0.2 eV in all cases, which implies that occupation of the tetrahedral sites is more favored for Ga regardless of the Ga concentration. Similar value of the energy gain (0.21 eV) was, also, theoretically obtained in[21] for moving precisely one Ga atom from octahedral to tetrahedral site, but calculated in the "dilute limit" for 80- and 160- atoms in YAG ($Y_3Al_5O_{12}$) supercell.

Despite $Ga^{3+}$ having a larger ionic radius than $Al^{3+}$, gallium clearly prefers to occupy tetrahedral sites which have smaller available volume than octahedral sites. Therefore, the differences in the formation of the chemical bond by Al and Ga should be taken into account in order to understand this apparent paradox. An analysis of the electronic structure obtained for our compounds during the calculations shows that both Al and Ga form the chemical bond in a different way depending on their environment. It can be well illustrated on the example of $Lu_3Ga_5O_{12}$ with only two inequivalent gallium atoms and one oxygen atom in its structure. Thus, in addition to shorter Ga(IV) – O distances (1.87504 Å for Ga(IV)-O and 2.01336 Å for Ga(VI)-O), the "center of gravity" of Ga(IV) valence electrons energies is shifted to stronger binding energies compared to Ga(VI). Together with the values of calculated ionic charge equal +1.8 for Ga(IV) and +1.9 for Ga(VI) this implies that Ga(IV) atoms form more covalent bond with surrounding oxygen atoms than Ga(VI). The same principles of bond formation hold for Al atoms, too. According to our calculations performed for mixed compounds, both Ga and Al cations being placed in the tetrahedral environment form more covalent bonds with the surrounding oxygen atoms as compared with the cations placed in the octahedral environment. It should be noted here, that Nakatsuka et al.[17] denote the same tendency in bond formation for $Y_3Al_{5-x}Ga_xO_{12}$ mixed compounds "from an estimate of the proportion of covalent bonding based on bond strength". Higher covalence of the chemical bond makes it energetically more favorable for both Ga and Al to occupy, if possible, the tetrahedral positions in the garnet structure. But, being more electronegative, Ga forms even more covalent bonds with the surrounding oxygen ions than Al, which makes it energetically more favorable to occupy the tetrahedral positions instead of Al.

The reason of formation of more covalent bond lies in the electronic structures of Ga ($[Ar]3d^{10}4s^24p^1$) and Al ($[Ne] 3s^23p^1$) atoms. Covalent bond has a highly directional character and its formation is possible when both cation and anion are able to form appropriately directed orbitals without substantial energy loss. The $4s^24p^1$ valence electrons of Ga and $3s^23p^1$ of Al are both suitable to form such directed orbitals through the hybridization of the s- and p- orbitals, resulting in $sp^3$-like hybridized orbitals with tetrahedral geometry. Such hybridized orbitals have a localized, directional character, while other types of hybridization suitable for formation of the bonds with six surrounding oxygens atoms would have a delocalized character proper for an ionic bond. Covalent bond being energetically more favorable, in addition, implies smaller positive charge on the cation, and Ga(IV), being more electronegative than Al(IV), has yet lower charge than Al(IV), which results in a lower cation-cation interaction in the Ga case. Moreover, Ga 3d electrons which are located in the valence band are not screened well from the outer charges. Being (almost) not involved in the formation of the chemical bond, they nevertheless experience the Coulomb interaction with the negative charges on the oxygen atoms, which reduces their binding energy. In the case of tetrahedral environment it is easier for Ga to adopt its five d- orbitals to minimize such interaction with only four negative charges located around, which can play an additional role in Ga occupation preferences. The differences in the formation of the chemical bonds between oxygen anions and various cations located in the tetrahedral and octahedral environment play role in the cation occupation preferences also for other garnets with other compositions. For instance, for Si, having [Ne] $3s^2 3p^2$ electronic structure and formal valence charge 4+ in $X_3Y_2Si_3O_{12}$ garnets (where $Si^{4+}$ ions occupy tetrahedral sites, only), it is energetically very favorable to form the chemical bond through the formation of four $sp^3$ hybridized orbitals directed to the oxygen atoms of a tetrahedron. Otherwise, in the case of Fe having $[Ar]3d^6 4s^2$ electronic structure, the delocalized character of its both s- and d- orbitals, almost removes any preferences for the occupation of the tetrahedral or octahedral positions.

The preferences for Ga to occupy the tetrahedral positions for any Ga content, as it follows from our calculations and the discussion above, is, nevertheless, not fully reflected in the results of our NMR experiments where some small fraction of Ga atoms is also found in the octahedral sites and the Ga(IV)/Ga(VI) ratio decreases with increasing Ga concentration. However, this can be explained in the following way: DFT calculations correspond to the ground state at 0 K, while in reality the cation distribution arises during the synthesis of the crystal at high temperatures (~ 2000 K). The energy difference of about 0.2 eV per Ga-Al pair is comparable to the thermal energy $k_BT$ (where T is the temperature of the garnet sample preparation or thermal treatment). Thus, the final cationic arrangement cannot be expected to be in an equilibrium occupations corresponding to 0 K, but rather to be shifted towards a random distribution of the cations – in dependence on the energy gains for the respective structures, and also in dependence on conditions of the sample preparation (thermal treatment). The fact that temperature treatment (high temperature annealing) can influence the cation distribution over tetrahedral and octahedral sites was demonstrated in Ref. 37. Our calculations show, that the energy gain per one Ga(tetra)+Al(octa) pair depend on Ga concentration (Table 8), which is in a good agreement with our experimentally obtained dependence of Ga(IV)/Ga(VI) ratio on *x* in $Lu_3Al_{5-x}Ga_xO_{12}$ solid solutions (see Fig. 10).

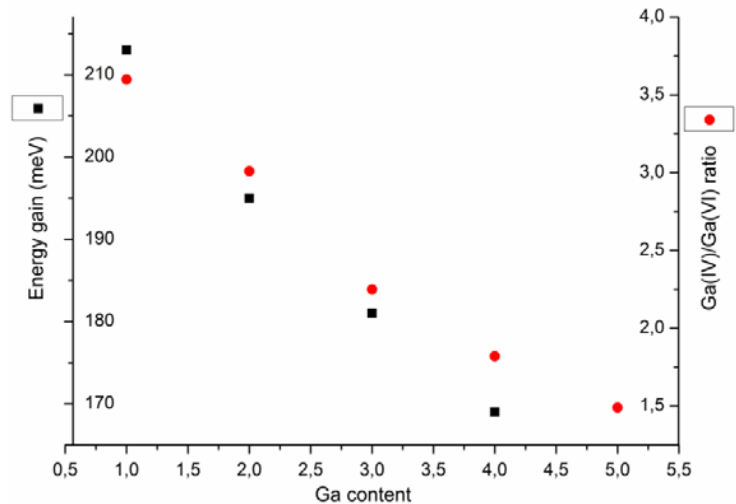

Fig. 10. Calculated energy difference of Ga placing into octahedron instead of tetrahedron (and vice versa for Al) in dependence on gallium content $x$ (black squares) and experimentally measured Ga(IV)/Ga(VI) ratio (red circles).

## 4. Summary

$^{27}$Al and $^{71}$Ga solid-state NMR measurements were carried out in the mixed $Lu_3Al_{5-x}Ga_xO_{12}$ compounds in the form of bulk and grinded single crystals. Single crystal, powder static and MAS spectra were measured and analyzed. This allowed us to determine full set of spectral parameters (chemical shifts, quadrupole frequencies) which describe the NMR spectra including the distribution of the Al and Ga ions over the tetrahedral and octahedral sites of the garnet lattice as a function of the Ga content x. In particular, it was found that both the $^{27}$Al and $^{71}$Ga quadrupole frequencies at tetrahedral sites practically do not vary with the Ga content in the solid solutions suggesting that the tetrahedrons do not undergo additional distortion. Contrary, quadrupole frequencies of both nuclei substantially increase in the octahedral sites in spite of the growth of octahedrons volumes with increase of Ga content, which cannot be explained in simple charge point model of EFG calculation. But such property is well reproduced in DFT calculation of the EFG tensor. Precision values of Al and Ga substitution numbers (Tables 3 and 5) were calculated from $^{71}$Ga and $^{27}$Al NMR spectra. We found that (Fig. 7) partial substitution of Al by Ga results in heterogeneous distribution of Ga atoms over the tetrahedral and octahedral sites in the garnet structure. Contrary to its higher ionic radius, Ga tends to occupy the tetrahedral positions which have a lower volume as compared to the octahedral ones. This is further supported by DFT calculations performed for all investigated compounds which show that, for a given Ga content in the garnet structure, allocation of Ga into the tetrahedral sites is energetically more favorable, regardless of Ga concentration. The energy difference between Ga(IV)+Al(VI) and Ga(VI)+Al(IV) scenarios is about 0.2 eV for one such Ga+Al pair, and this value slightly decrease with increase of Ga content.

The Ga occupation preferences were explained by the differences in the electronegativity values of Al and Ga atoms, and by the different nature of the chemical bond formation in the tetrahedral and octahedral environment. Both Al and Ga being in the tetrahedral environment have a larger covalent component of the chemical bond compared to the octahedral environment, but higher value of electronegativity of Ga atom makes such a difference higher for Ga atoms and makes it energetically more favorable to substitute Al(IV). Larger covalence of the chemical bond in the case of the tetrahedrons was explained by the symmetry of valence orbitals of these atoms. Because in both cases the chemical bond is formed, mainly, by the cation's s- and p- valence electrons, the hybridized orbitals will have localized character with tetrahedral geometry, favorable for the formation of directed covalent bond. Otherwise, six surrounding oxygen atoms belonging to the octahedrons would cause the formation of more ionic bond with delocalized character.

Partial substitutions of $Al^{3+}$ ions by larger $Ga^{3+}$, accompanied by an increase of the lattice constant, result in substantial deformation of both Al and Ga octahedrons, leaving the geometry of both tetrahedrons relatively unchanged, but change the orientation of their axial axes.

**Acknowledgements:** The financial support of the Czech Science Foundation grant 17-09933S is gratefully acknowledged. Computational resources were provided by the CESNET LM2015042 and the CERIT Scientific Cloud LM2015085, provided under the program "Projects of Large Research, Development, and Innovations Infrastructures".